%% file: bare_conf.tex
\PassOptionsToPackage{pdfpagelabels=false,bookmarks=false,hidelinks}{hyperref} 
\documentclass[10pt, conference, compsocconf]{IEEEtran}

\usepackage{wrapfig}
\newcommand*\circled[1]{\tikz[baseline=(char.base)]{
          \node[shape=circle,draw,fill=black,text=white,font=\bf,inner sep=0.5pt] (char)
            {\scriptsize#1};}}
\usepackage{subcaption}
\usepackage{graphicx}
\usepackage{fancyhdr}
\usepackage{color}
\let\labelindent\relax
\usepackage{enumitem}
\usepackage{tikz}
\usepackage[font={small},textfont={bf}]{caption}
\usepackage[labelfont=bf]{caption}
\usepackage{comment}
\graphicspath{{images/}}
\usepackage[T1]{fontenc}

\usepackage{amsmath}
\usepackage{hyperref}
\usepackage{mathtools}
\usepackage{float}
\usepackage{cite}

\def\footnoterule{\kern-3pt
  \hrule \kern 2.6pt} 

\let\OLDthebibliography\thebibliography
\renewcommand\thebibliography[1]{
  \OLDthebibliography{#1}
  \setlength{\parskip}{0pt}
  \setlength{\itemsep}{0pt plus 0.3ex}
}

\begin{document}
\title{Multiverse: Dynamic VM Provisioning for Virtualized High Performance Computing Clusters} 

\author{\IEEEauthorblockN{Jashwant Raj Gunasekaran\IEEEauthorrefmark{1}, Michael Cui\IEEEauthorrefmark{2}, Prashanth Thinakaran\IEEEauthorrefmark{1}, Josh Simons\IEEEauthorrefmark{2}, Mahmut T. Kandemir\IEEEauthorrefmark{1}, Chita R. Das\IEEEauthorrefmark{1}}
\IEEEauthorblockA{\IEEEauthorrefmark{1} Computer Science and Engineering, The Pennsylvania State University, \IEEEauthorrefmark{2} VMware Inc.} 
\{jashwant, prashanth, mtk2, das\}@cse.psu.edu,
\{xiaolongc, simons\}@vmware.com\vspace{-2mm}}
\maketitle

\bstctlcite{IEEEexample:BSTcontrol}
\
 \pagenumbering{gobble}


\pagestyle{plain}

\vspace{-3mm}
\input{0-abstract}
\vspace{-1mm}
\begin{IEEEkeywords}
\hspace{0.in} HPC; virtualization; VM provisioning; cloning;
\end{IEEEkeywords}

\vspace{-2mm}
\section{INTRODUCTION} 
\label{sec:intro}
\input{1-introduction}
\section{BACKGROUND AND MOTIVATION} 
\label{sec:background}
\input{2-background}
\section{Overall Design of \textit{Multiverse}}

\label{sec:modeling}
\input{3-modeling}
\section{Implementation Methodology}
\label{sec:scheme}
\input{4-scheme.tex}

\section{Experimental Setup}
\label{sec:experiment}
\input{5-methodology}
\section{Evaluation and Results} 
\label{sec:results}
\input{6-results}

\section{Related Work}

\label{sec:related}
\input{7-related}

\section{Conclusion and Future Work } 
\label{sec:conclusion}
\input{8-conclusion}

\section*{Acknowledgments}
\label{sec:acknowledgment}
\input{9-acknowledgment.tex}

{\scriptsize \bibliographystyle{IEEEtran}
\bibliography{IEEEabrv,references}}

\end{document}

%% file: 0-abstract.tex

\begin{abstract}
Traditionally, HPC workloads have been deployed in bare-metal clusters; but the advances in virtualization have led the pathway for these workloads to be deployed in virtualized clusters. However, HPC cluster administrators/providers still  face challenges in terms of resource elasticity and virtual machine (VM) provisioning at large-scale, due to the lack of coordination between a traditional HPC scheduler and the VM hypervisor (resource management layer). This lack of interaction leads to low cluster utilization and job completion throughput. Furthermore, the VM provisioning delays directly impact the overall performance of jobs in the cluster. Hence, there is a need for effectively provisioning virtualized HPC clusters, which can best-utilize the physical hardware with minimal provisioning overheads.

Towards this, we propose {\em Multiverse}, a VM provisioning framework, which can dynamically spawn VMs for incoming jobs in a virtualized HPC cluster, by integrating the HPC scheduler along with VM resource manager. We have implemented this framework on the \textit{Slurm} scheduler along with the \textit{vSphere} VM resource manager. In order to reduce the VM provisioning overheads, we use instant cloning which shares both the disk and memory with the parent VM, when compared to full VM cloning which has to boot-up a new VM from scratch. Measurements with real-world HPC workloads demonstrate that, instant cloning is 2.5$\times$ faster than full cloning in terms of VM provisioning time. Further, it improves resource utilization by up to 40\%, and cluster throughput by up to 1.5$\times$, when compared to full clone for bursty job arrival scenarios. 
\end{abstract}


%% file: 1-introduction.tex
\vspace{-1mm}High performance computing (HPC) has evolved over the years, due to the application demands ranging from scientific computing to AI/ML-based applications~\cite{tsai2018overview}. HPC system stack is changing rapidly to keep up with the performance demands of such applications. It is an important constraint for the HPC cluster administrators to improve the cluster throughput and utilization, without sacrificing the application performance. Therefore, in the past these HPC applications have been traditionally deployed on bare-metal hardware~\cite{rad2015benchmarking}, due to the performance guarantees offered by native execution. 

On the other hand, with the advent of high throughput computing jobs~\cite{htc}, HPC clusters require massive scaling of resources while accessing large volumes of data. Towards this, there have been several advancements for rapid HPC provisioning such as OpenStack~\cite{openstack}, which enables software-defined HPC infrastructure, that saves the time wasted on manual configuration and cluster provisioning. {This leads to better infrastructure manageability~\cite{Kureshi2013}, while ensuring the native bare-metal performance. However, this performance guarantee comes at the cost of abysmal bare-metal cluster utilization, due to lack of fine-grain support in sharing resource like CPU, memory and network across multiple tenants~\cite{Huang:2006:CHP:1183401.1183421}.}


Recent advancements in hypervisor and virtualization technology can improve the cluster utilization without sacrificing the performance while supporting multi-tenant execution. Through virtualization, HPC workloads can also benefit from supporting resource heterogeneity, performance isolation, improved security, etc.,~\cite{hwang2015netvm}. Moreover, the performance overheads of virtualized HPC clusters with respect to job execution times, have {less than 5\% overheads for throughput and non-I/O intensive applications}, when compared to bare-metal clusters~\cite{virtual-hpc,hpc-ml}. Combined with these numerous benefits, nowadays virtualization of HPC environments is becoming more prevalent~\cite{ali2011performance,dynamic-lb}. For instance, the Johns Hopkins University Applied Physics Laboratory, transformed their existing bare-metal cluster to a virtual cluster (vGRID)~\cite{vmware-josh}, which led to drastic improvements in resource utilization by upto 19\%, with less than 4\% performance degradation. 

Although, virtualization guarantees near-native job execution time, private virtualized HPC clusters still face challenges in terms of cluster management such as, (i) efficient scalability of the VMs in the cluster, (ii) efficient cluster utilization  with respect to dynamic job arrivals at the HPC scheduler, and (iii) minimizing the VM provisioning delays. This is because, the native HPC job schedulers like \textit{Slurm~\cite{slurm} or \textit{Torque}~\cite{torque} do not interact with the hypervisor management layer}~\cite{7379288,hassan2016scalability}. Therefore, HPC cluster administrators/providers need to statically provision and manage Virtual Machines (VMs) for every incoming job from the HPC scheduler. Motivated by these observations, we argue that, there is a need for a framework to dynamically provision virtualized HPC clusters, which ensures efficient cluster utilization, with minimal provisioning overheads.

To overcome some of the challenges mentioned above, several frameworks have been proposed to integrate traditional HPC schedulers ~\cite{slurm,torque}, along with VM resource managers~\cite{openstack,1opennebula}, to enable seamless and self-managed VM provisioning for HPC clusters. Still, a majority of these frameworks lack support to dynamically provision VM for every job and requires the cluster administrator to manually intervene during scale-out phase~\footnote{Scale-out indicates that, we have more incoming jobs than available VMs.} and mapping the job requirements from the HPC scheduler to the VMs. Furthermore, in an attempt to reduce the VM provisioning delays, several optimizations such as image sharing, taking snapshots of VM~\cite{li2017towards,zaslavsky2015creating}, etc,. have been proposed. Despite these efforts, new VMs still take hundreds of seconds for provisioning \cite{formosa3,vslurm}. There have been recent efforts to reduce VM provisioning latencies using {\em instant clone technology~\cite{lagar2009snowflock}}. Instant clones uses a copy-on-write~\cite{bhimani2017accelerating} feature to drastically reduce the provisioning time when compared to full clones~\cite{cloning}, which has to boot-up a new VM from scratch. To the best of our knowledge, there are very few existing works, which have support for dynamic VM provisioning ~\cite{meier2016dynamic, vslurm,formosa3}, however, they do not utilize techniques such as instant clone to reduce provisioning overheads in a virtualized HPC environment.
 
To holistically address the shortcomings of existing works, we build a dynamic VM provisioning framework, called \textit{Multiverse}, which can spawn new VMs for incoming jobs using instant cloning in a virtualized HPC cluster. This enables more flexible and cluster utilization-aware VM provisioning. In summary, we make the following {\bf contributions}:
\begin{itemize}[leftmargin=*]
\item \textcolor{black}{We have designed \textit{Multiverse}, a generic framework that {\em integrates} HPC scheduler with VM orchestrator\footnote{We use the terms orchestrator and resource manager interchangeably. } for dynamic VM provisioning on a per-job basis}. \textcolor{black}{We have implemented a prototype of the mentioned framework using Slurm~\cite{slurm} as the HPC job scheduler and VMware vSphere ~\cite{vsphere} as the VM orchestrator}. 
\item We incorporate an admission control system and a dynamic load balancer using sqlite3 database~\cite{sqlite3}, which ensures efficient VM placement decisions in cluster.
\item We have characterized the performance and scalability of the \textit{Multiverse} framework for two types of VM cloning mechanisms on a 220 core HPC cluster. 
\item Our experimental analysis using different job arrival scenarios shows that, instant cloning is $2.5\times$ - $7.2\times$  better than full cloning, in terms of VM provisioning time. Further, our results show that instant cloning can provide up to 40\% better resource utilization, and  $1.5\times$ better cluster throughput, when compared to full cloning.
\end{itemize}








%% file: 2-background.tex


\subsection{Why Virtualization for HPC?} 
Despite the fact that virtualization having proved to be cost effective, scalable and reliable in majority enterprise infrastructures, HPC applications are still executed on bare-metal, non-virtualized clusters (for most cases), to achieve maximum performance. This suffers from major challenges such as (i) lack of support for {dynamic} load balancing and migration, and (ii) lack of isolation and security among multi-tenant workloads. However, virtualization can transform HPC infrastructure by overcoming these challenges by enabling (i) proactive VM migrations during failure and load imbalance, (ii) micro-segmentation using network virtualization for fine-grain isolation.\vspace{5mm}  
\begin{figure}[hbpt]
\centering
    \includegraphics[width=0.45\textwidth]{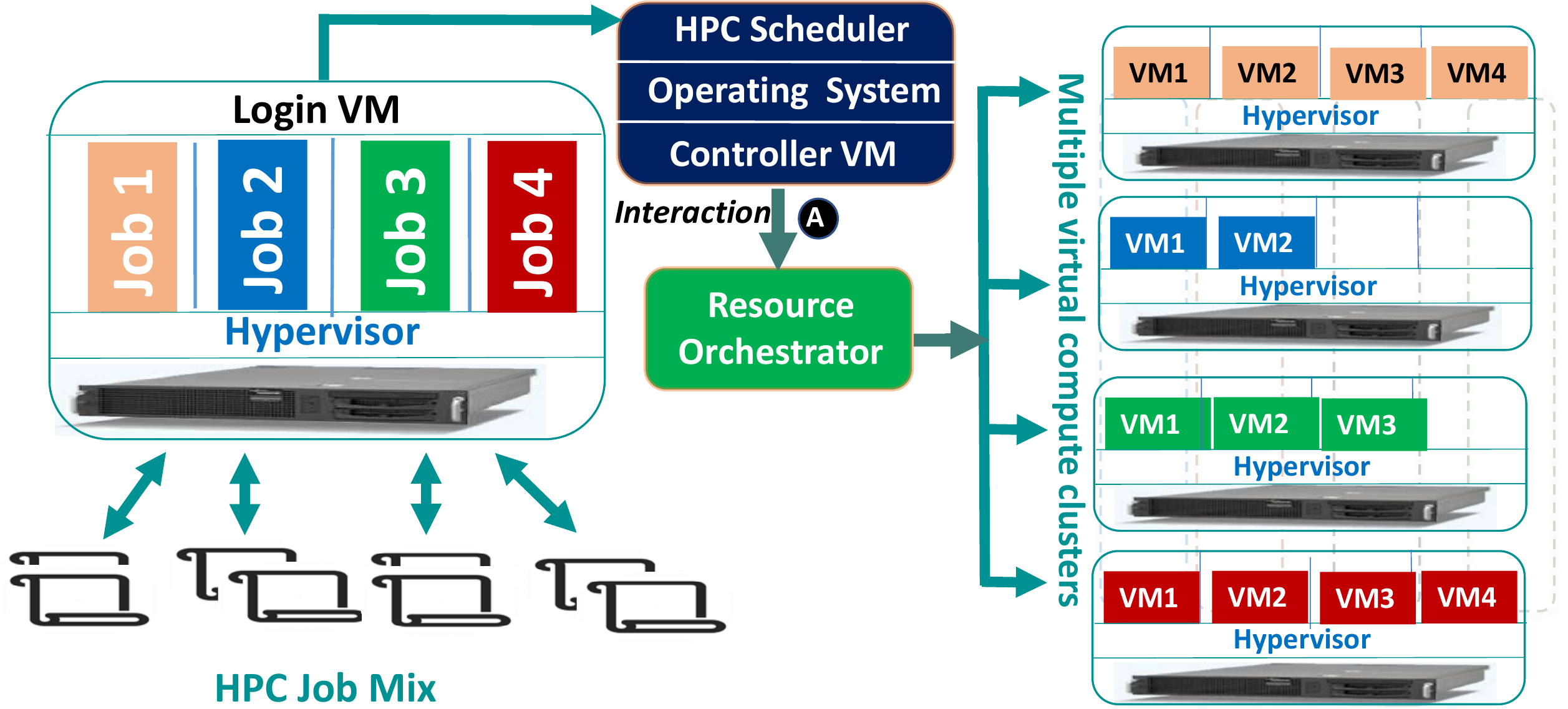}
    \caption{Overview of a Virtualized HPC framework.}
    \label{virtual}
\end{figure}

\subsection{Overview of Virtualized HPC framework}
Figure~\ref{virtual} illustrates the architecture of virtualized HPC framework. All physical nodes in the cluster are virtualized using a VM hypervisor (VMware {ESXi} or kvm \cite{kvm}). The hypervisor which directly runs on the physical nodes in privileged mode provides abstraction for VMs to run while mapping host resources such as CPU, memory, storage, and network to each VM. The {original login node, master node for job scheduling, and compute nodes now run as VMs on the existing login, controller, and compute nodes, respectively, within the same cluster. In addition, a VM orchestrator ({VMware vCenter} or Opennebula)~\cite{vsphere,1opennebula} runs inside a VM on one controller node and provides centralized management of the hosts and VMs, coordinating resources for the entire cluster}.  
In this paper, we propose a dynamic VM provisioning {framework for allocating VM(s)} for every incoming job in such a virtualized HPC cluster.

\subsection{Challenges in Dynamic VM Provisioning}
In case of a large HPC cluster with hundreds to thousands of nodes, where many jobs can arrive within a short span of time, cluster administrators face fundamental challenges to statically allocate and manage VMs for those jobs. Dynamic provisioning would be better in such scenarios and also lead to better resource utilization, because it avoids over-provisioning of VMs. Typically, VM provisioning is handled by resource managers such as VMware vSphere, OpenStack or KVM~\cite{vsphere,1opennebula,kvm}. However, traditional HPC schedulers do {\em not} have support to provision and manage VMs (as shown in Figure~\ref{virtual}~\circled{A}). Therefore, there is a need to integrate HPC schedulers to work collectively with VM resource managers and make such integration transparent to the user. We explain the challenges in such a design and how we achieve this integration in Section~\ref{sec:scheme}. 

\subsection{VM Cloning Types}
Fast and agile dynamic provisioning of VMs is directly impacted by the time taken to provision a new VM. Typically, virtualized environments use cloning~\cite{cloning}, to provision new VMs. A clone is a copy of an existing VM, which is called the parent of the clone. There are two well known types of clone (i) {\em full clone}, which is an independent copy of a virtual machine that shares nothing with the parent virtual machine after the cloning operation, and (ii) {\em linked clone}, which is a copy of a virtual machine that shares virtual disks with the parent virtual machine.

Recent interest in container-based resource provisioning has led to the developments of a new cloning technique called {\em instant clone}~\cite{lagar2009snowflock}. Instant clone uses rapid in-memory cloning of a running parent VM, and leverages copy-on-write to rapidly deploy VMs. They are much faster when compared to full and linked clones but a significant amount of time needs to be spent to configure and customize the network. We explore the opportunity to use instant clone and compare its performance with full clones for different inter-arrival times of jobs. The results of the characterization are presented and discussed in Section~\ref{sec:results}.   


%% file: 3-modeling.tex

In this section, we first discuss the challenges in our proposed design to integrate HPC schedulers with VM orchestrators. Subsequently, we explain the in detail the design of \textit{Multiverse} framework that effectively addresses the challenges.
\subsection {Design Challenges}
It is not trivial to enable the integration of HPC schedulers along with VM resource managers due to the following reasons. First, we need to hide the virtualization integration from the user. Second, multiple jobs might share the same parent VM/image while cloning new VMs. This leads to potential issues related to disk management (snapshots) and concurrency in cloning. Third, VMs have to be customized as specified in the job requirements file submitted to the HPC scheduler. This requires support for availability of multiple different VM images {when using instant clone}. Finally, the resource orchestrator needs to have policies for dynamic load balancing and admission control of VMs. Typically this would be handled by the HPC scheduler for all jobs, since we override the scheduler functionalities to support integration with VM resource orchestrators, it is the responsibility of resource orchestrator to handle load balancing and admission control.  

\subsection{Design Choices} 
We explain the design of \textit{Multiverse} with respect to overcoming the challenges mentioned in the previous section. First, to retain the original job submission behaviour of users, we need to customize the job submission workflow in the HPC scheduler. Typically, jobs go through different stages within the scheduler starting from job submission to resource selection  to job allocation. We modify each of these stages as follows. During the job submission, we parse all the job requirements so that, later we can {\em dynamically provision} a new VM based on the requirements. In the resource selection stage, we make the job wait until the VM for the job has been provisioned and, finally, in the job allocation stage, we ensure that the job is allocated to the corresponding VM. 

Second, to support launching VMs for concurrent job submissions, our system needs to be thread safe.\begin{wrapfigure}{r}{0.15\textwidth}
    \centering
    \includegraphics[width=.15\textwidth]{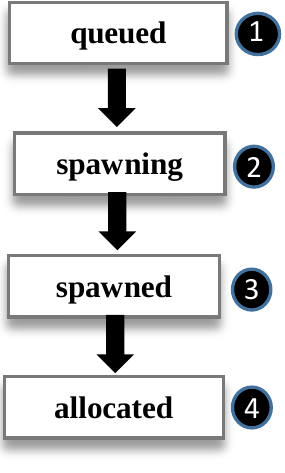}
        \caption{State machine used in our \textit{Multiverse} design.}
\label{states}
\end{wrapfigure} Though, the schedulers themselves are  thread safe for multiple job submissions, the changes we make to every stage in the scheduler should ensure the same. Our design makes use of an {\em explicit state machine}, which is thread safe, for maintaining the different states of a job, as shown in Figure~\ref{states}. Once a job is submitted, it enters into queued state~\circled{1}, and continues to exist in that state until a VM spawn\footnote{We use the terms launch and spawn interchangeably.} process is initiated for that job. Then, it moves to spawning state~\circled{2}, while the VM is being spawned and configured. Once the spawning is complete, the job moves to a spawned state~\circled{3}. Next the job should be scheduled on the newly-spawned VM. However, the scheduler is unaware about which VM to allocate the job,  because there could be multiple VMs with similar configuration that can satisfy for a job. Hence, we need a mechanism to enforce the scheduler to allocate a job on to a specific VM, which was spawned for the job. To do so, we make use of job-feature parameters (predominantly supported in all HPC schedulers) within every job and VM, to enable a unique {job-to-VM} mapping. After the job is allocated on the VM, it moves to the allocated state~\circled{4}. We explain in detail about the implementation of the state machine in Section~\ref{sec:scheme}.

Apart from being thread-safe, we also need to ensure that, multiple VM clones from the same parent image does not add additional overheads on disk management. This in turn leads to clone failures. To mitigate such clone failures, we make use of a {\em rate-limiter mechanism}, which can limit the number of clones per parent VM for a given time. Through our characterization study, we set the rate-limiter at 15 clones per minute and 200 clones per second for full clone and instant clone, respectively. Note that instant clone supports more concurrent clones as it shares both disk and memory with the parent VM.

Third, to support different customizations of the VM in accordance to the job requirements, we provide a baseline VM image (based on the OS used) which is used for cloning of VMs. Any application-specific libraries which might be required by the users can be made available on a shared file systems (like NFS) and this file system is mounted on the cloned VMs. We also have to enable the scheduler specific configurations such that, the new VM will be added to the scheduler's node pool. This is done by using customization scripts which are executed on the VM right after cloning. 

Fourth, to enable admission control and load balancing, we {design} an \textit{utilization aggregator} to store physical node-specific metrics like CPU, memory utilization, number of active VMs, etc,. This database can be queried by the resource orchestrator using custom APIs, to check current resource utilization against specific admission control policies. Also, by leveraging the resource utilization metrics from the database, we design two {\em load balancing policies}, to enable fair resource allocation in the cluster. The detailed implementation of the system is given in Section~\ref{sec:load}. 
\begin{figure}
\begin{minipage}[t]{0.95\linewidth}
    \centering
    \includegraphics[width=0.99\linewidth]{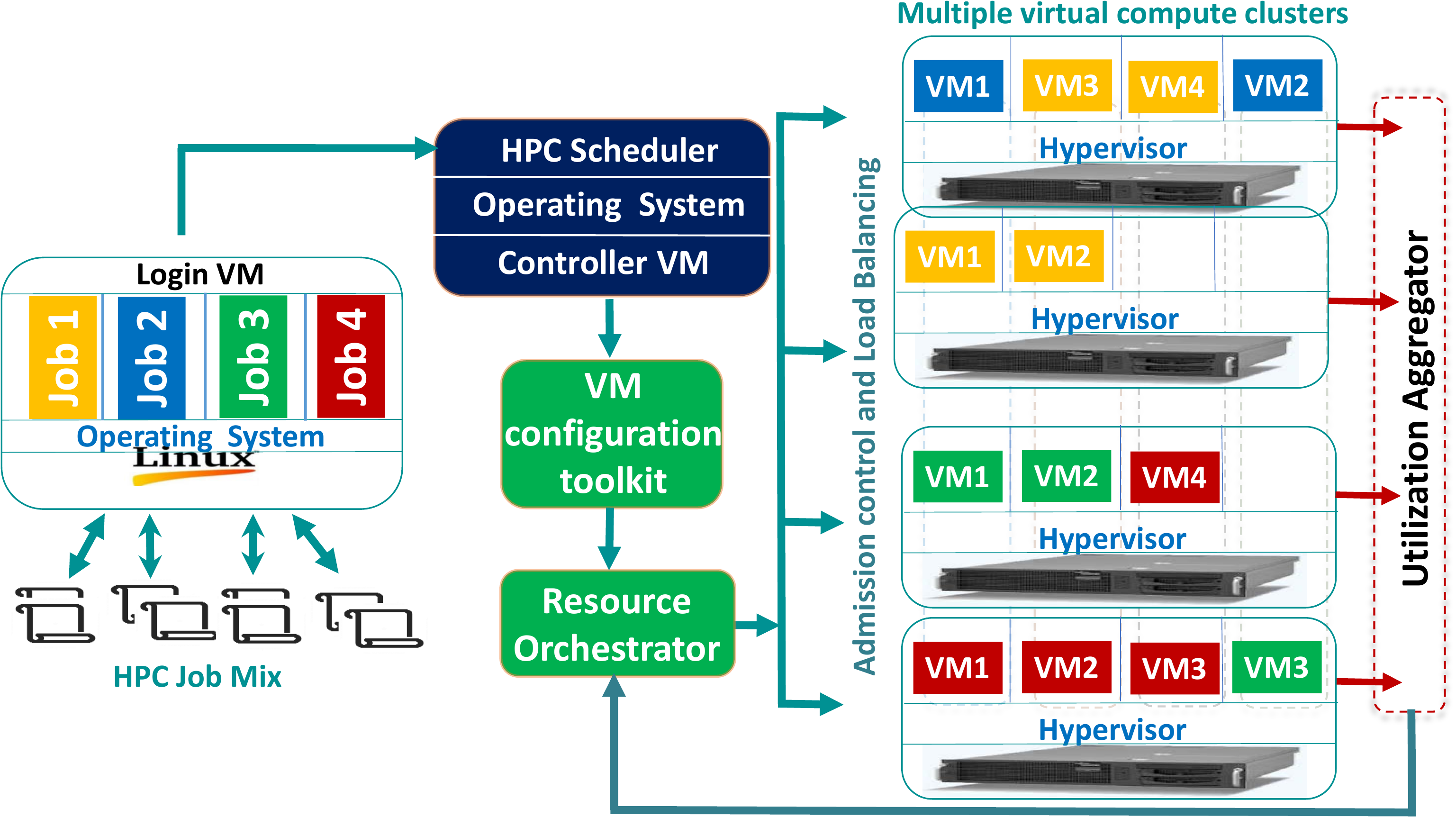}
        \caption{A high level overview of \textit{Multiverse} framework. The VM configuration toolkit acts as an interface between scheduler and resource manager. Utilization aggregator (shown in red) sends real-time cluster utilization metrics to the VM resource manager (shown in green).\vspace{-3mm}}
        \label{multiverse}
\end{minipage}
\end{figure}
\subsection{\textit{Multiverse} workflow}
The overall workflow of the \textit{Multiverse} framework is shown in Figure~\ref{multiverse}. Users submit a job to a login node, which is extracted by the controller node. The controller node decides job scheduling and placement based on specified policies. As briefly discussed in the previous subsection, we override this functionality of the controller, to spawn a VM for every job submitted and allocate the job on the spawned VM. To achieve this design, we make use of custom scheduler plugins to extract the job requirements, spawn a new VM based on the requirements, allocate the job and delete the VM after job completion. For all the VM specific interactions, we need to use a VM configuration toolkit which can interface with the resource orchestrator. The resource allocation and management of the VMs on to physical nodes is handled by the resource orchestrator/manager by leveraging the cluster specific metrics, which are exposed to a database using our \textit{utilization aggregator}.

%% file: 4-scheme.tex
\begin{figure}
    \centering
    \includegraphics[width=0.45\textwidth]{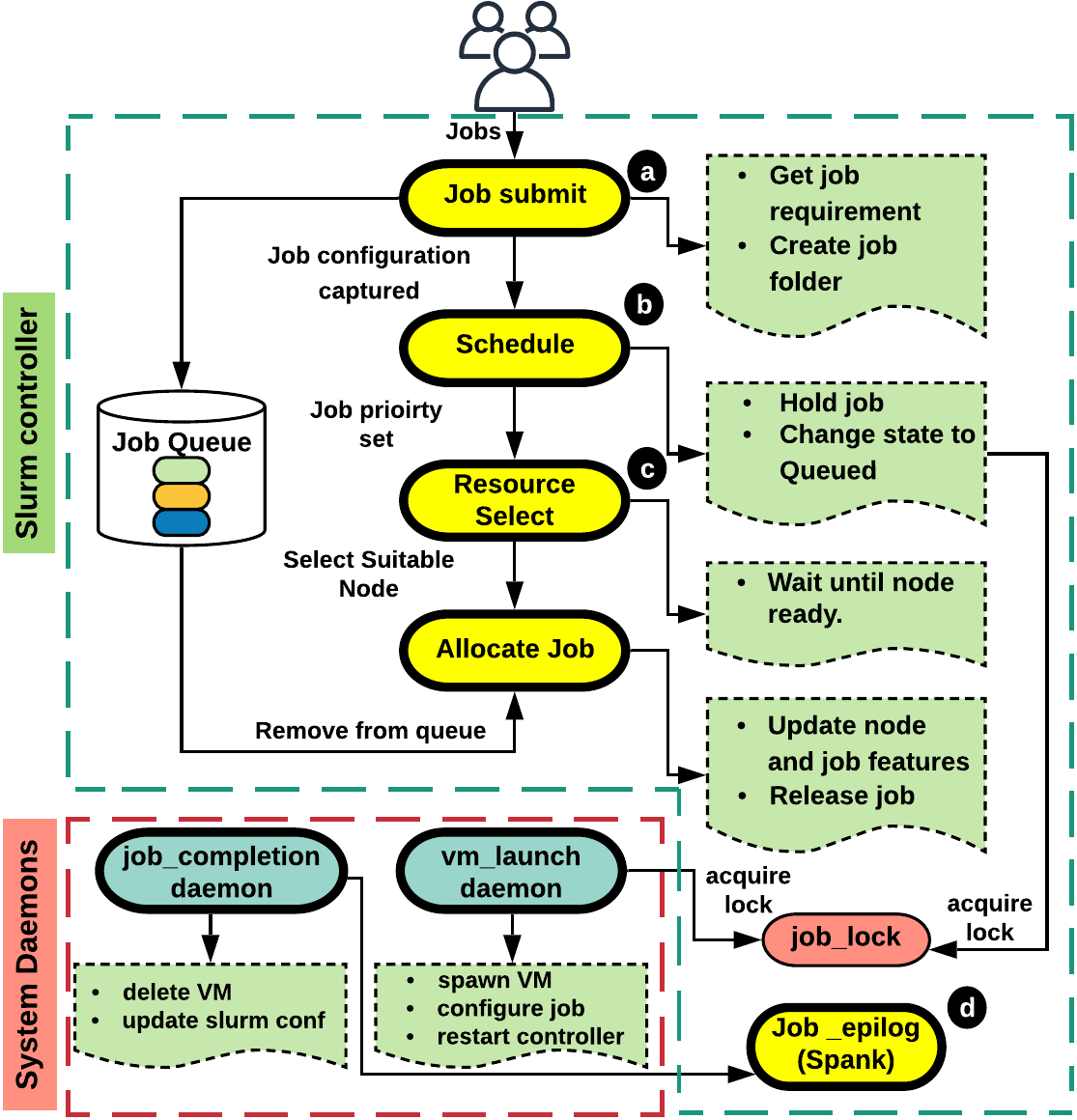}
        \caption{Plugin-based implementation of \textit{Multiverse} framework, showing the various steps of interaction. The system daemons are shown in blue and the \textit{Slurm} plugins are shown in yellow. Job lock (shown in red) is shared between the plugins and daemons.\vspace{-3mm}}
\label{fig:plugin}
\end{figure}
We explain in detail about the implementation of our proposed \textit{Multiverse} framework. While our implementation is with respect to \textit{Slurm} scheduler and \emph{vSphere} resource manager, \textit{Multiverse} is generic to be extended to work with most HPC schedulers and VM resource managers.
\subsection{Augmenting \textit{Slurm}}
As shown in Figure~\ref{fig:plugin}, we make use of four custom \textit{Slurm} plugins to enable support for dynamic VM provisioning. The details of these plugins are given below.
\subsubsection{\textbf{Job Submit Plugin}} Job submit~\circled{a} is called by \textit{Slurm} controller right after submitting the job and before scheduling resource allocation. This plugin can override the existing \texttt{job\_submit} method, to give user-defined controls and change the configuration parameters of the job. We use this plugin to create  a job config file and copy the following information to it: job name, number of CPUs, required memory, minimum number of nodes, submit time and other job related metrics. The job config file has a uniquely identifiable name, which is a concatenation of job name with submission timestamp. 

\subsubsection{\textbf{Scheduler plugin}}Scheduler plugin~\circled{b} is called after the~\texttt{job\_submit} and before resource (VM/node) selector. We override the~\texttt{slurm\_sched\_p\_initial\_priority} function using scheduler plugin, which sets the initial priority for the job. We set the priority value such that, all incoming jobs will be on hold and will not be eligible to schedule (we name this as~\texttt{sched\_hold}). This is essential, because the VMs for allocating the jobs are spawned only after job submission. Also, in the same function, we update the job information (job name and \textit{Slurm} generated \texttt{job\_id}) to a file named \texttt{queued\_jobs}. We employ a locks to ensure atomic writes to the file by multiple jobs. 

\subsubsection{\textbf{Resource Select Plugin}} The resource select plugin~\circled{c} is invoked after the \texttt{job\_submit} and scheduler plugin to check if there are any available resources (VMs) to run the job. Since the VMs for the job are spawned after submission, they might not be ready during this phase of scheduling. Therefore, we modify this plugin to return true for VMs selection by default, though there are no available resources.  

\subsubsection{\textbf{Spank Plugin}} The spank plugin~\circled{d} is used to define any user-defined functions which is called during various steps in \textit{Slurm} execution depending upon the context during which it is called. 
We use this plugin in \texttt{job\_epilogue} context to call a cleanup function. This function will notify the controller node that the job has been complete. The state of compute VM (the newly spawned VM for executing the job) is marked as ``down" to prevent future jobs to be scheduled on this VM. Also, the job output and error logs are copied to master and login node.

\subsection{Custom system daemons}
Apart from customizing the \textit{Slurm} plugins, we also design and implement two custom system daemons which will handle the tasks of VM creation, job launch and VM deletions. They are described in detail below.\footnote{The circled annotations in each subsection are with respect to Figure~\ref{states}.}

\subsubsection{\textbf{VM launch daemon}}
The primary purpose of this daemon, is to initiate a VM launch for all submitted jobs to \textit{Slurm}. The daemon is designed to work like a state machine (as shown in Fig~\ref{states}) where every job can be in one of the following four states. 
\begin {itemize}[leftmargin=*]
\item	Queued~\circled{1}: The job is added to \textit{Slurm} queue and is updated in queued\_jobs file by the  scheduler plugin. For jobs in this state, the daemon calls the corresponding function  to start spawning a new VM as per the job requirements which were captured using the job\_submit plugin. The new state of the job is changed to spawning. 

\item Spawning~\circled{2}: The daemon calls vm\_launch script for a previously queued job. Now the job will be periodically queried to see if VM spawning is complete. The daemon takes necessary actions (re-spawn or cancel) if the spawning fails due to some reason. 

\item Spawned: If the VM spawning for the job is complete, the daemon changes the state of the job to spawned. For all jobs in spawned state~\circled{3}, the daemon updates the \textit{slurm} config file to include details of the newly spawned VM. It releases the jobs from \textit{Slurm} hold so that the job can be allocated~\circled{4} to the corresponding VM for execution. Also, after adding any new nodes to \textit{slurm} config file, the \textit{Slurm} controller has to be restarted. This is due to the inherent design of \textit{Slurm}. Hence, the daemon also restarts the controller, for the VM allocation to be successful for all spawned jobs. 

\item Pending: Since we use locks to ensure serialized write to the queued\_job file, the job\_lock has to be acquired by the \textit{Slurm} scheduler plugin and the VM\_launch daemon. Hence, in the \textit{Slurm} scheduler plugin, if the job\_lock is busy, it updates the job information to a pending\_job file. The daemon constantly extracts jobs from this file and initiates the VM\_launch function. Hence the pending state, is  an auxiliary state used when the job\_lock is busy. 

\end{itemize}

\subsubsection{\textbf{Job completion daemon}}
This daemon, constantly monitors the state of all the compute VMs. Recall from our spank \texttt{job\_epilog} plugin, the state of VMs after job completion is marked to be ``down". For all such VMs, the job completion daemon calls a cleanup function which will ensure the following two steps. First, it clears the node information from \textit{Slurm} config file. Second, it deletes all the job configuration details captured during submission and also deletes the VM which was spawned for running the job. 

\subsection{Admission control and load balancing}
\label{sec:load}
We developed a python-based api which can query the real-time information about all hosts in a cluster and maintain the information in a \textit{sqlite}~\cite{sqlite3} database. We use this api in the VM\_launch function to get a compatible host for cloning new VMs. This is a convenient API which can be extended to develop different admission control and load balancing schemes. The API exposes basic functionalities such as (i) initializing a database with existing cluster information, (ii) update the database based on new allocations/de-allocations, and (iii) get a compatible host for the new clone request Using these functionalities of the API, we designed an admission control and load balancing policy for the \textit{Multiverse} framework.
\subsubsection{Admission control} 
We enforce two types of admission control.
If all the resources of the hosts are currently utilized or there is not enough room to accommodate a new request, the job waits in the queue until resources become available. If the required resources of the job are more than the physical capacity of the host, the job is revoked from execution. In the former case, to avoid starvation of the jobs due to unavailability of resources, we make sure that newly incoming jobs are queued behind the delayed job inside the internal queued\_jobs file used by vhpc\_launch daemon. There is still scope to improvise this policy by enforcing rules for starvation. For example we can set a limit on how many times the job can be re-queued, or how long the scheduler can run other smaller jobs until the bigger jobs keep waiting for resources.
\subsubsection{Load balancing}
We have two different policies for load balancing the VMs across the hosts. In the first policy, we chose the first compatible host by doing a linear search across all hosts. In this context, "compatible" means that the host has {\em enough resources} to be allocated for the VM requirements. However, there can be more than one compatible host in the cluster. Hence we implement a second policy, where we randomly select {a} host from the list of compatible hosts. This incurs additional overhead compared to {the} first\_available policy but can ensure better load balancing across the cluster.  
                                          
\subsection{Software specifications}
\subsubsection{Scheduler and APIs}We use \textit{Slurm} version 19.05 as our HPC scheduler. All the \textit{Slurm} plugins are written in \texttt{C-language} and generated as shared library (.so) files, which are dynamically linked to the scheduler. The system daemons are \texttt{bash} files which are hosted as \texttt{systemctl} service in {Linux}. We also design our VM\_launch and cleanup scripts as bash files. For implementing \texttt{mutex} locks we make use of \texttt{Flock}~\cite{flock} Linux utility which is available in both bash and C.  vHPC toolkit~\cite{vhpc} was developed using Python which makes use of \texttt{pyVmoi}~\cite{pyvim} and \texttt{pyVim} python packages that enable access to \textit{vSphere} APIs. We use this toolkit for VM configuration based on job requirements. 
\subsubsection{Clone configuration}
We need to generate a clone configuration file for spawning a new VM using vHPC toolkit. For full clone, the template VM can reside in any node in the cluster. But in the case of instant clone, we cannot instantiate clones on different hosts, other than the template VM. Therefore we have a template VM on every node of the cluster and based on the chosen host given by our load balancer, we initiate the instant clone on that host.

In addition, for instant clone the CPU and memory cannot be dynamically configured , because it uses VMFork~\cite{lagar2009snowflock} to fork off a new VM from the template. Essentially, the same hardware configuration of the template VM would be retained for the cloned VM. In the case of diverse jobs which have different memory and CPU requirements, we can have different-sized template VMs on each host and select a closest matching compatible template VM. 
\subsection{Assumptions and Limitations}
\label{sec:overheads}
As soon as a user submits a job to slurm, the job would be waiting to get scheduled until a VM is spawned for the job. Hence, the additional time incurred to clone new VMs is accounted along with the total waiting time of the user. However the cloning time would not affect other running jobs because it is implemented as background process that can be executed in parallel with other processes. Due to the inherent design of slurm, each time a new VM is added to the configuration file, we have to restart the \emph{Slurm} controller daemon. This overhead can be avoided if we use other HPC schedulers like PBS~\cite{pbs} or Torque~\cite{torque}, wherein new nodes can be added online without restarting controller daemons.  

Our proposed design changes to \emph{Slurm} might affect its inherent job scheduling features. For instance, \emph{Slurm} supports different scheduling policies like backfill, priority etc. Since the scheduling policies now depends on the VM scheduler used by \emph{vSphere}, administrators can use the corresponding scheduling policies in \emph{vSphere}, which reflect the same behaviour as \emph{Slurm}. Moreover, the original scheduling policies of the HPC scheduler can be retained if we do not tag every job to its respective VM. Its easier to customize this change into other HPC schedulers, when compared to \emph{Slurm}.


%% file: 5-methodology.tex
\subsection{Hardware Configurations}
\subsubsection{Cluster}We use a  {cluster of five Dell PowerEdge R630 nodes}, with 44 cores and 256 {GB} memory in each node \textcolor. The cluster has a 72 TB shared datastore with 5 physical network adapter of 10Gbps for each node. We use {ESXi}-6.7 hypervisor along with vSphere-6 resource manager.
\subsubsection{Master node} \textit{Slurm} controller is initiated from a master VM which is preallocated on the cluster. It runs on Centos 7 with 8 {vCPUs and 16 GB} memory which is large enough to handle our input job sequence.
\subsubsection{Login node} Users submit jobs via login node which is also a VM configured with 2 {vCPUs and 4 GB} memory. 


\subsection{Workload Generator}

We generate a workload for a job-sequence modeled after \textit{Poisson} inter-arrival times using a mean job arrival rate of $\lambda=10$ for 100 jobs. The jobs are associated with one of the three benchmarks which are (i) High-Performance Conjugate-Gradient (HPCG),a  simple additive Schwarz, symmetric Gauss–Seidel preconditioned conjugate-gradient solver\cite{hpcg}, (ii) High Performance Linpack (HPL)~\cite{hpl} which measures the performance solving a dense linear equation system, and (iii) MPIRandom-access~\cite{hpcc} which measures peak capacity of the memory subsystem while performing random updates to the system memory. 

We configure two types of jobs. First, we configure a short job which uses 2 vCPUs and 4 GB memory. Second, we configure a large job which uses 8 vCPUs and 16 GB memory. Both jobs only use 1GB of local storage (disk) space. We randomly sample from both the jobs for the sequence of 100 jobs. The HPCC and HPCG {input files} are modified accordingly to generate sufficient load for the given CPU and memory configuration. Each job has a running time ranging from 140s to 350s depending on the benchmark.  



%% file: 6-results.tex
\subsection{Evaluation Methodology}

We perform our evaluations from two complementary angles. First, we compare the overall job completion time in terms of cloning time, other overheads {(explained below)} and actual job running time for two types of cloning techniques. Second, we individually characterize every overhead incurred in our framework. The overheads and their description are shown in Table~\ref{tbl:overheads}.
\begin{wrapfigure}{l}{0.25\textwidth}
    \centering
    \includegraphics[width=.25\textwidth]{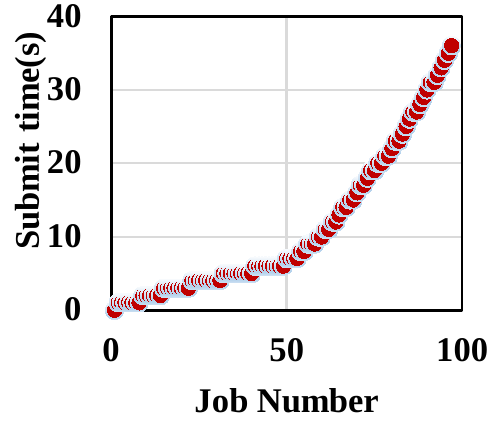}
        \caption{Poisson arrival rate fora sequence of  100 jobs with $\lambda = 10$.}
\label{arival}
\end{wrapfigure}
We evaluate two different job arrival scenarios as shown in Figure~\ref{arival}, (a) The first 50 jobs of the Poisson distribution, such that the cluster is fully utilized (workload-1) and (b) all the 100 jobs of the Poisson distribution, with 2x CPU over-commitment enabled in the cluster (workload-2). Further to characterize the bottleneck with full clone with respect to concurrent cloning, we also use a constant inter-arrival time of 10s, for jobs such that they all don't arrive within a short span of time.
\begin{figure*}[ht]
\begin{subfigure}[t]{.49\textwidth}
\centering
\includegraphics[width=\textwidth]{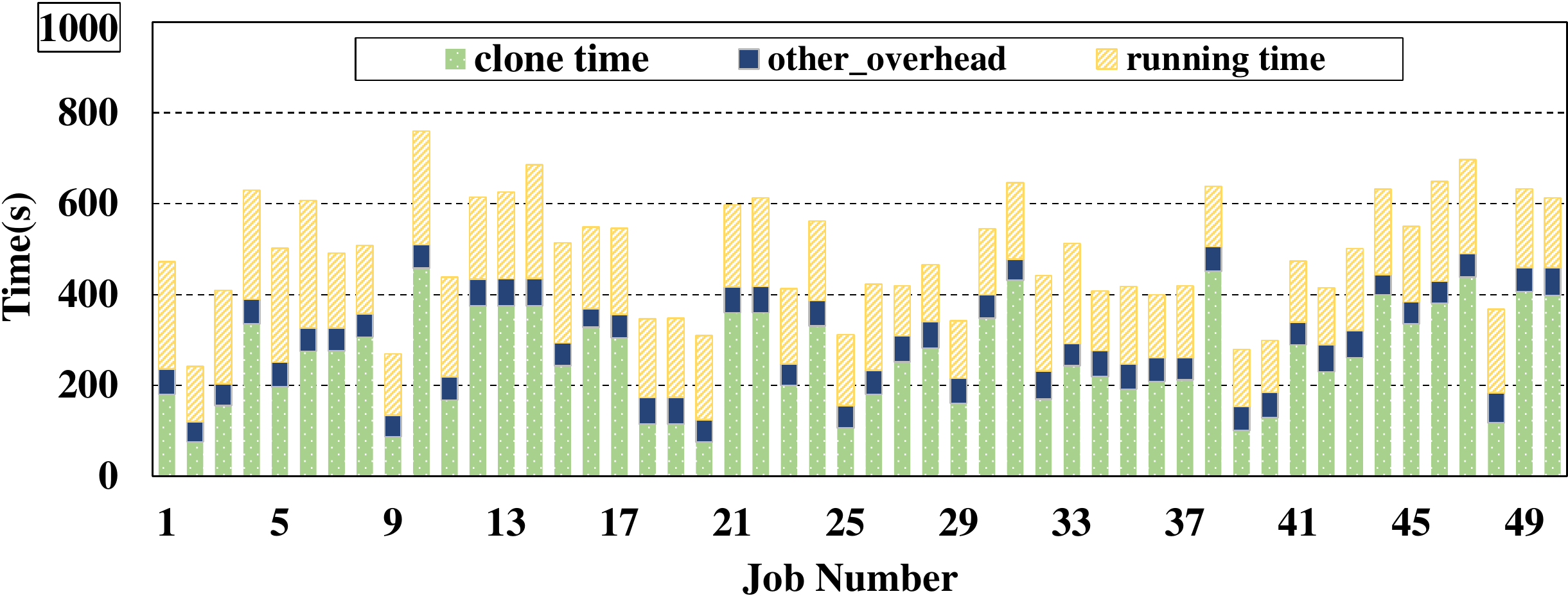}
\caption{Full clone.}
\label{mixed-full-50-c}
\end{subfigure}
\begin{subfigure}[t]{.49\textwidth}
\includegraphics[width=\textwidth]{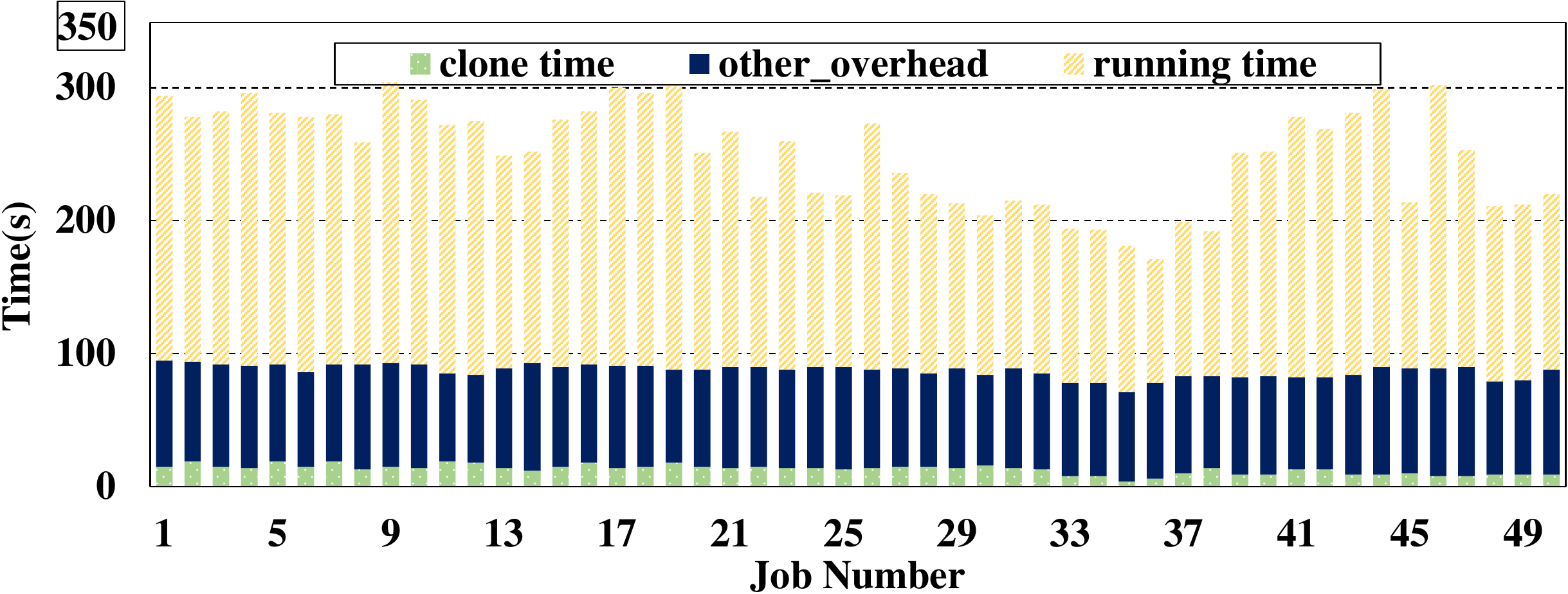}
\caption{Instant clone.}
\label{mixed-instant-50-c}
\end{subfigure}
\caption{Workload-1: Breakdown of Job Completion time in terms of cloning time, other overheads and job running time. }
\label{mixed-50-c}
\end{figure*}
\begin{figure*}[ht]
\begin{subfigure}[t]{.49\textwidth}
\centering
\includegraphics[width=\textwidth]{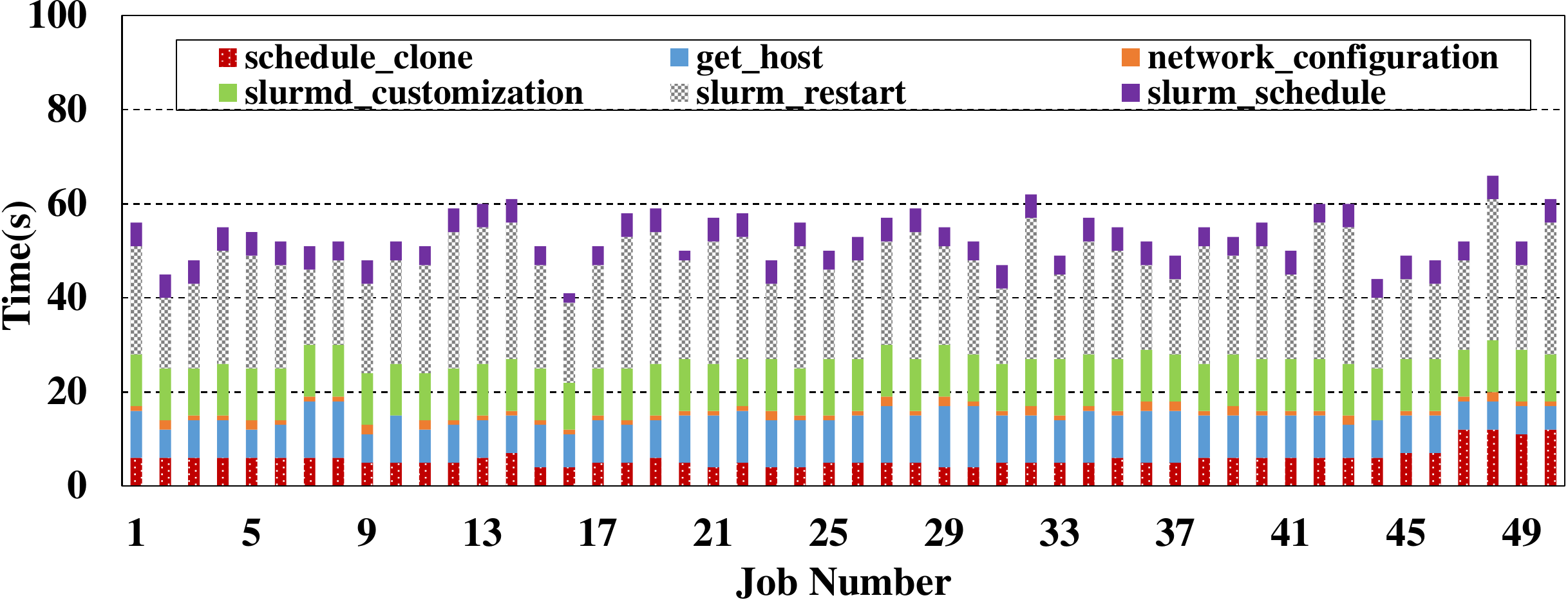}
\caption{Full clone.}
\label{mixed-full-50-b}
\end{subfigure}
\begin{subfigure}[t]{.49\textwidth}
\includegraphics[width=\textwidth]{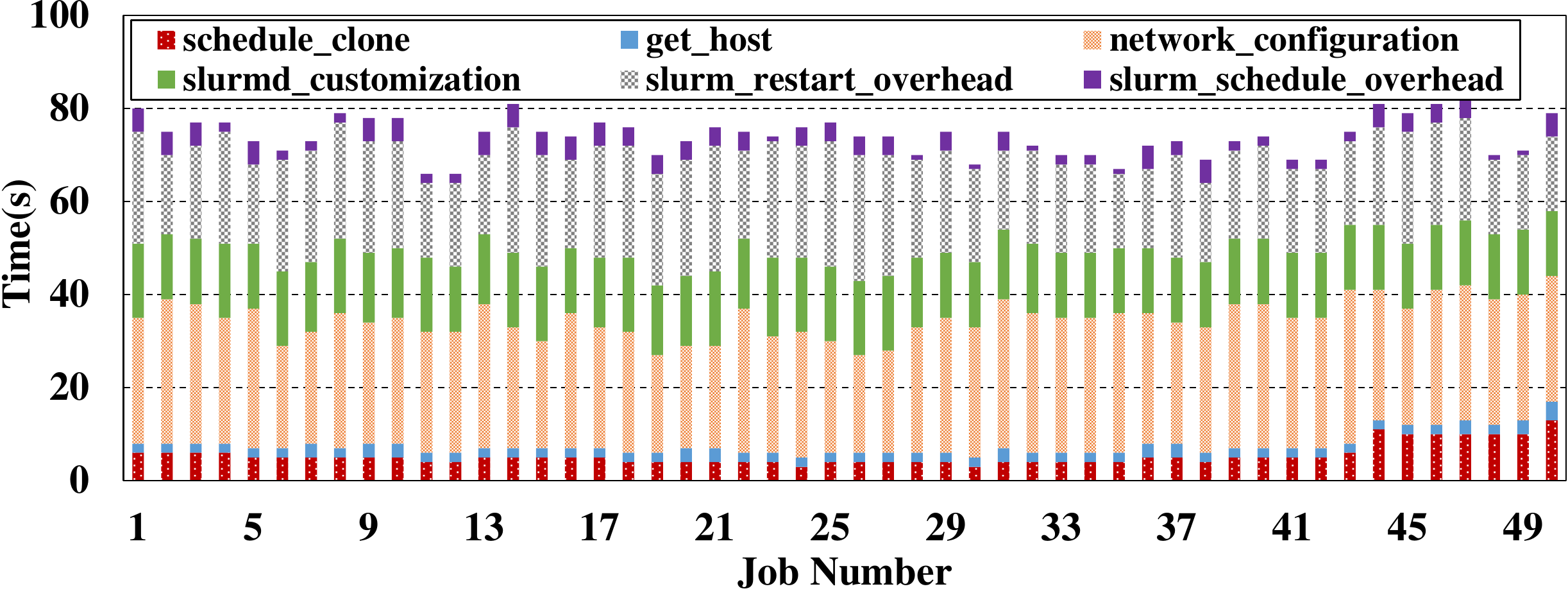}
\caption{Instant clone.}
\label{mixed-instant-50-b}
\end{subfigure}
\label{mixed-50-b}
\caption{Workload-1: Breakdown of other overheads for 50 jobs. }
\end{figure*}

\begin{figure*}[ht]
\begin{subfigure}[hbpt]{.5\textwidth}
\centering
\includegraphics[width=\textwidth]{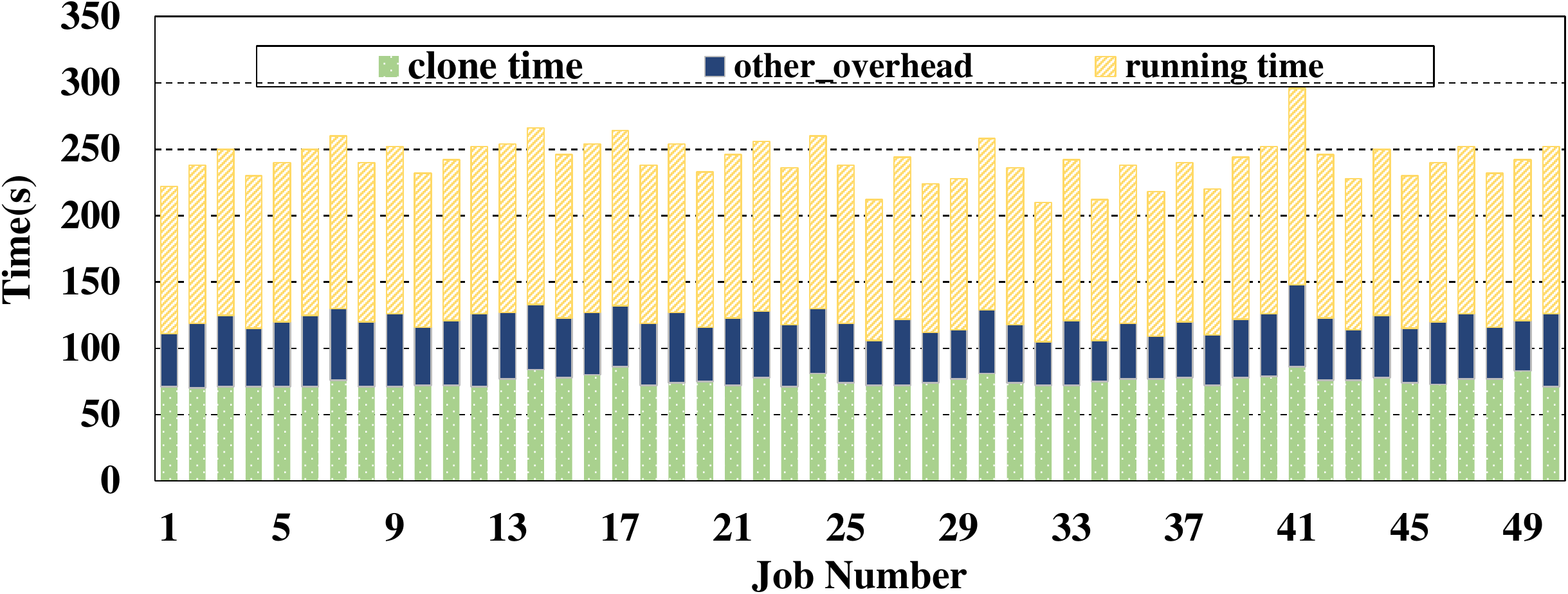}
\caption{Breakdown of Job Completion time in terms of cloning time, other overheads and job running time.}
\label{mixed-50-full-periodic-c}

\end{subfigure}
\begin{subfigure}[hbpt]{.5\textwidth}
\includegraphics[width=\textwidth]{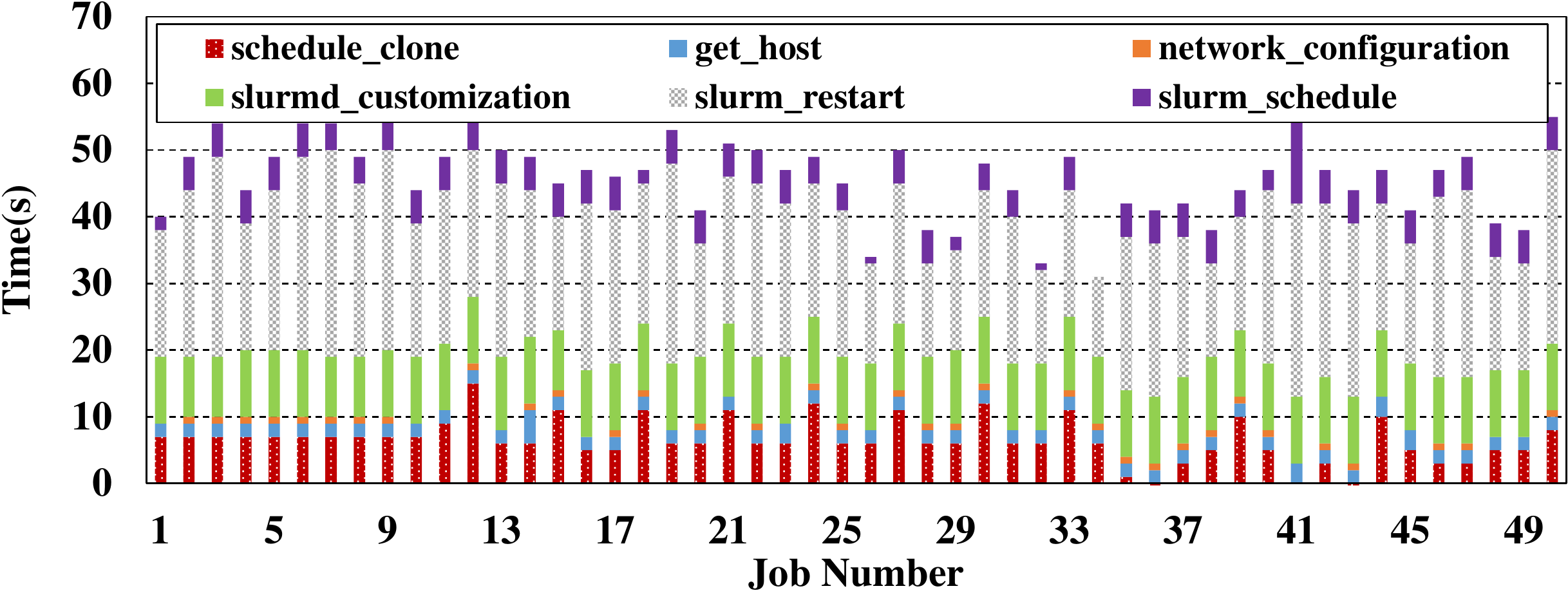}
\caption{Breakdown of other overheads.}
\label{mixed-50-full-periodic-b}
\end{subfigure}
\caption{Constant job arrival for 50 jobs using full clone.\vspace{5mm}}
\label{mixed-50-full-periodic}
\end{figure*}

\begin{figure}[ht]
\centering
\includegraphics[width=0.45\textwidth]{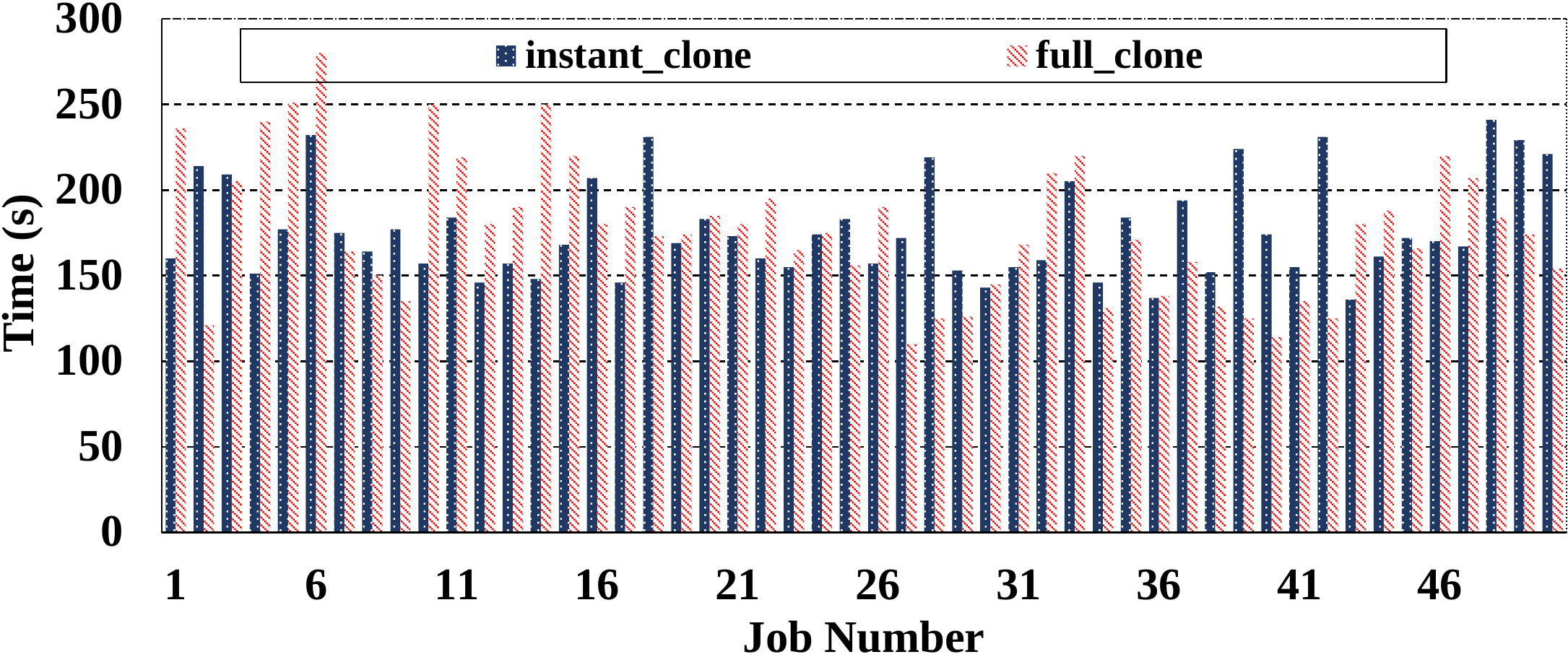}
\caption{Workload-1: Comparison of Job running time for full and instant clone for 50 jobs.}
\label{job_completion}
\end{figure}
\begin{figure*}[ht]
\begin{subfigure}[t]{.49\textwidth}
\centering
\includegraphics[width=\textwidth]{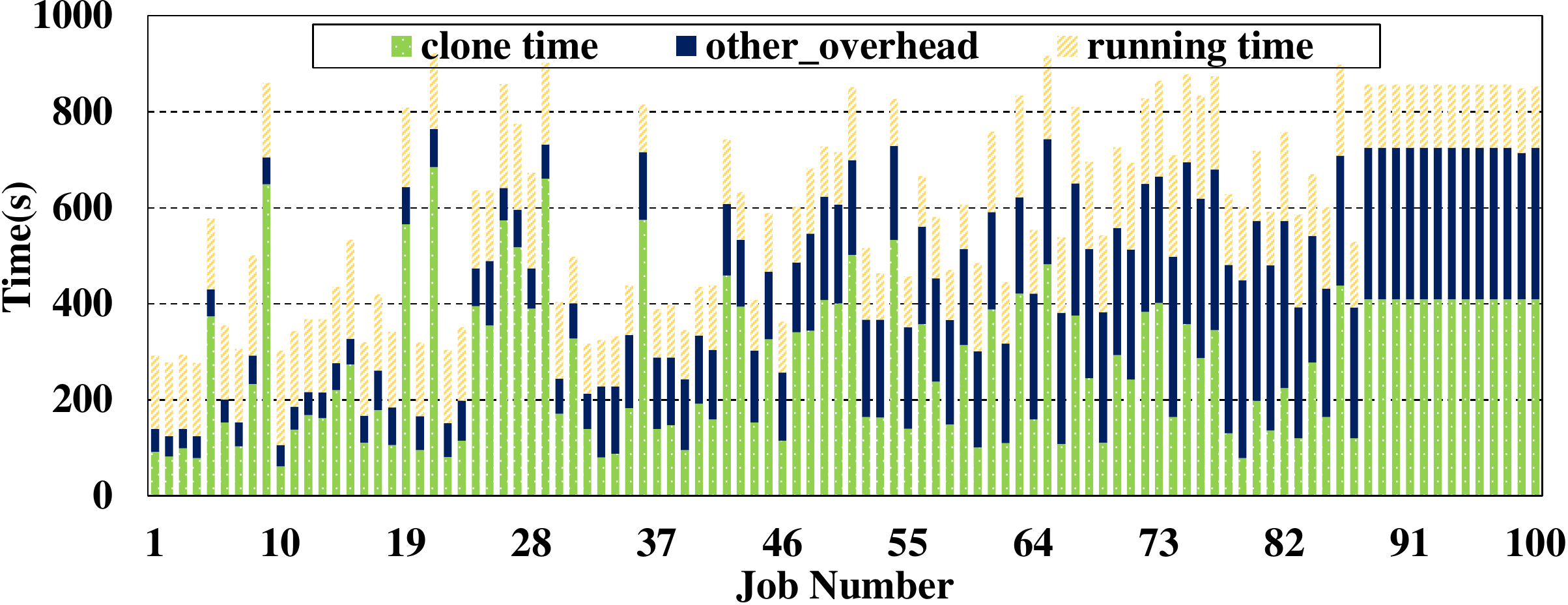}
\caption{Full clone.}
\label{mixed-full-c}
\end{subfigure}
\begin{subfigure}[t]{.49\textwidth}
\includegraphics[width=\textwidth]{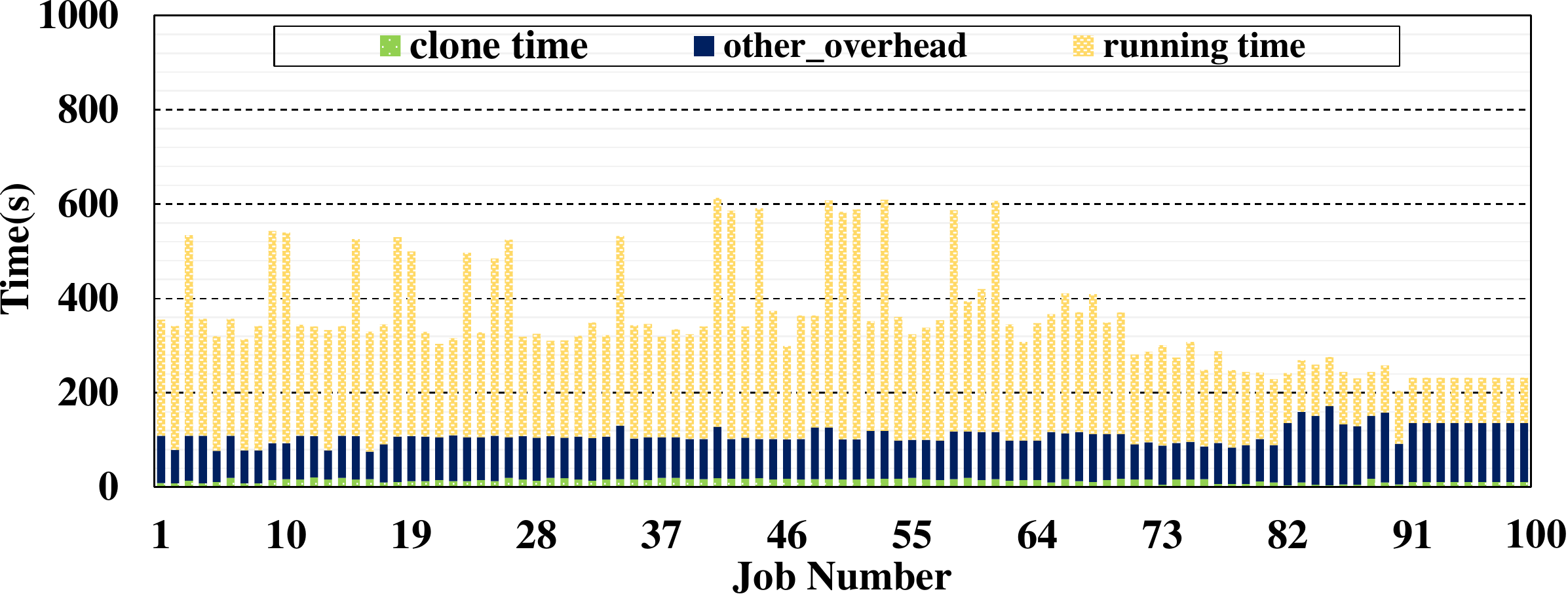}
\caption{Instant clone.}
\label{mixed-instant-c}
\end{subfigure}
\caption{Workload-2: Breakdown of Job Completion time in terms of cloning time, other overheads and job running time with 2x CPU over-commitment enabled in the cluster.}
\label{mixed-100-c}
\end{figure*}
\begin{figure*}[ht]
\begin{subfigure}[t]{.49\textwidth}
\centering
\includegraphics[width=\textwidth]{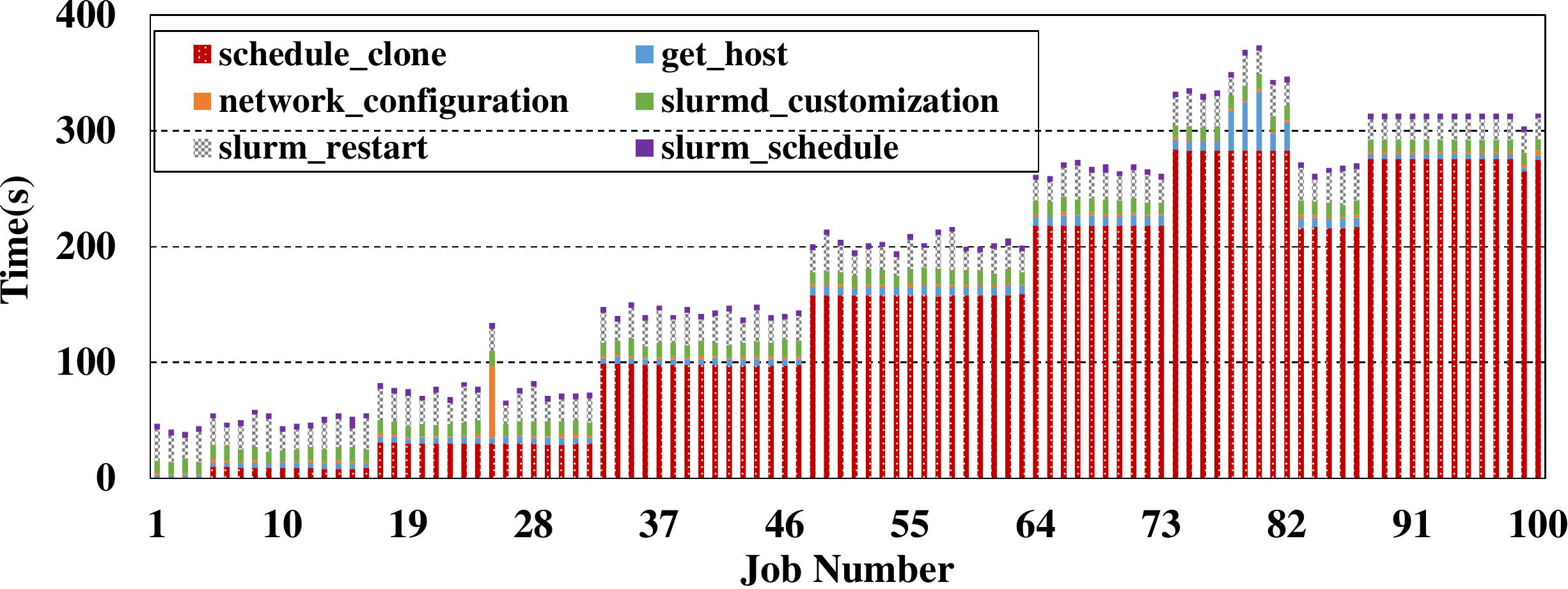}
\caption{Full clone.}
\label{mixed-full-b}
\end{subfigure}
\begin{subfigure}[t]{.49\textwidth}
\includegraphics[width=\textwidth]{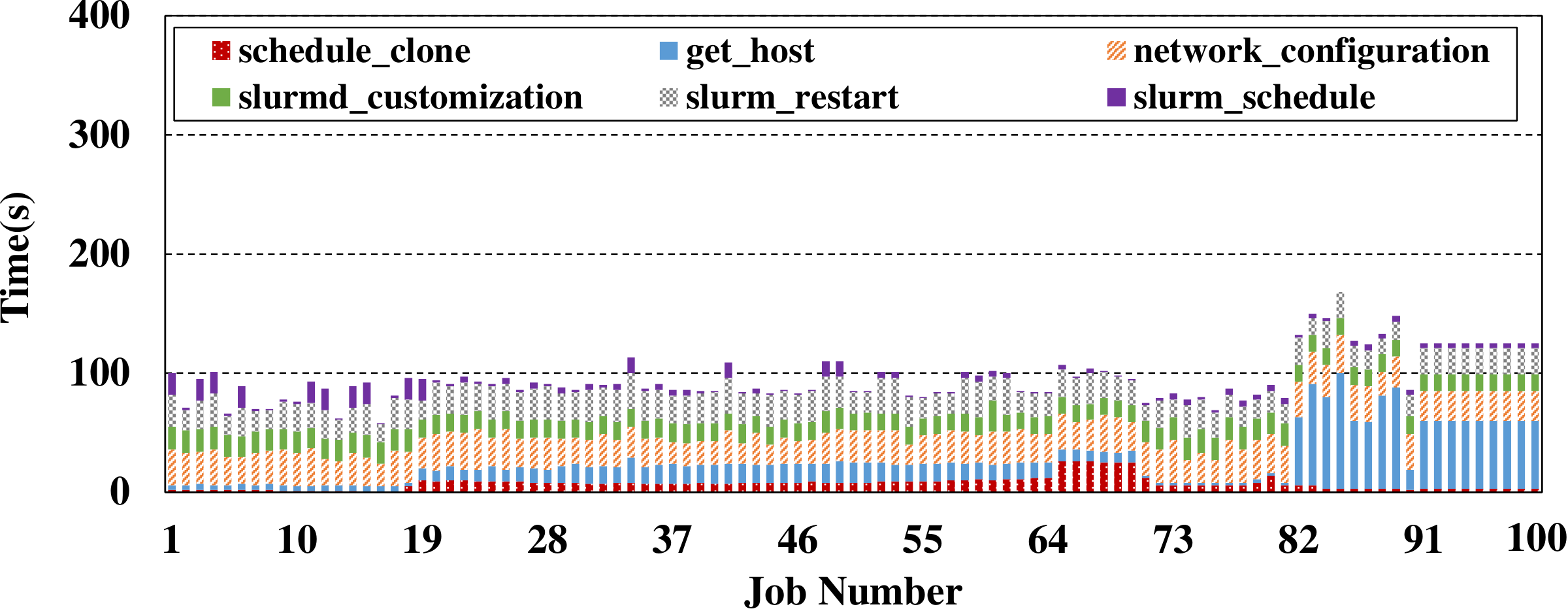}
\caption{Instant clone.}
\label{mixed-instant-b}
\end{subfigure}
\caption{Workload-2: Breakdown of other overheads with 2x CPU over-commitment enabled in the cluster for both clone types. }
\label{mixed-100-b}
\end{figure*}

\begin{figure*}[ht]
\begin{subfigure}[t]{.49\textwidth}
\centering
\includegraphics[width=\textwidth]{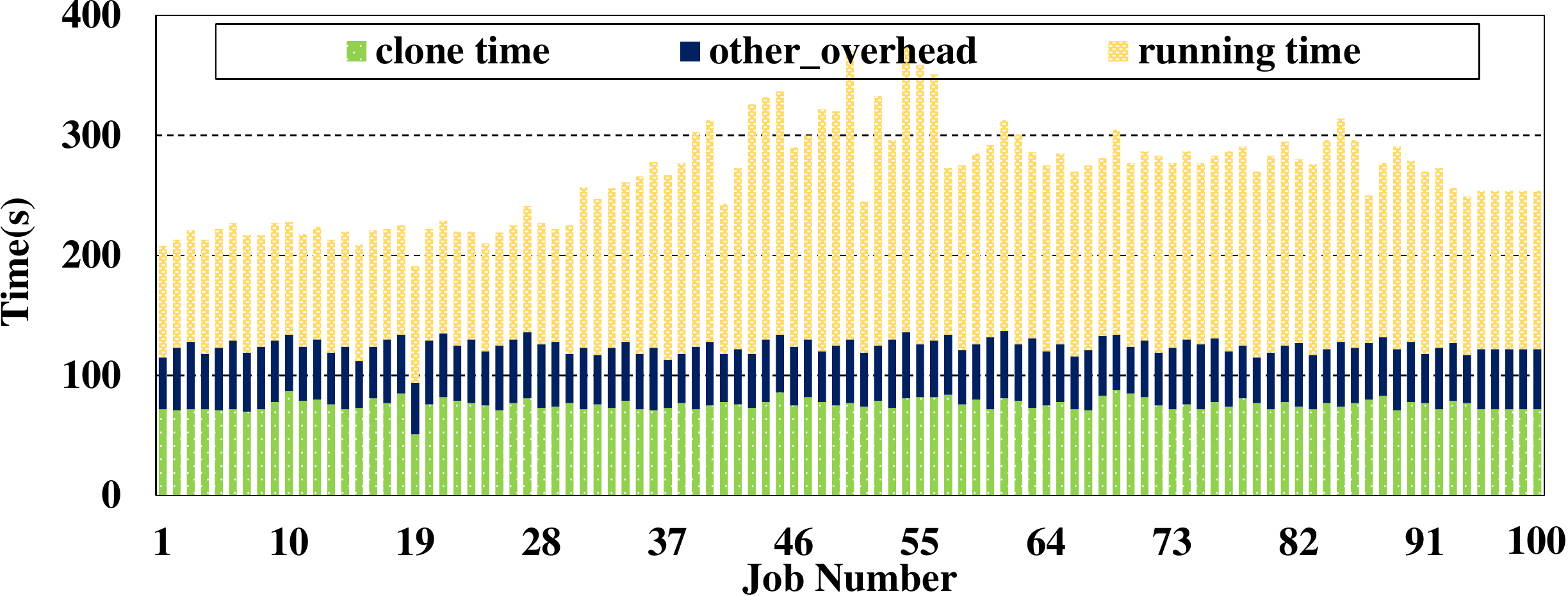}
\caption{Breakdown of Job Completion time in terms of cloning time, other overheads and job running time.}
\label{mixed-full-periodic-c}

\end{subfigure}
\begin{subfigure}[t]{.49\textwidth}
\includegraphics[width=\textwidth]{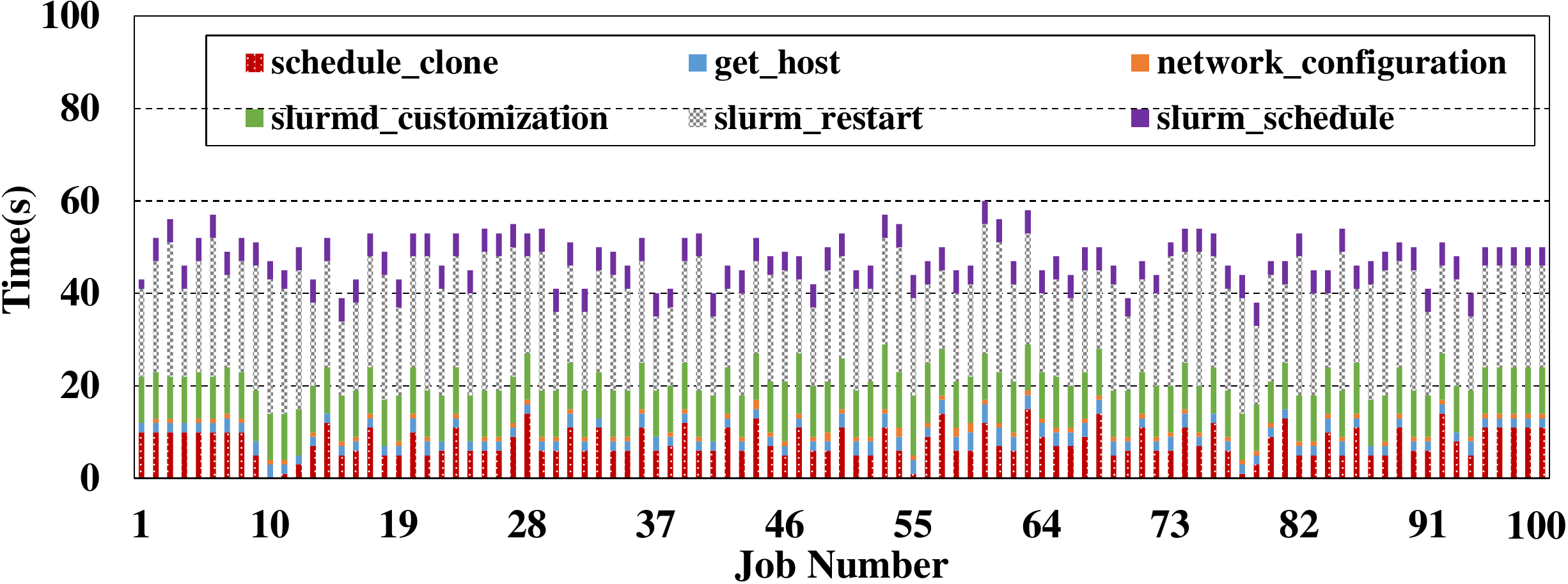}
\caption{Breakdown of other overheads.}
\label{mixed-full-periodic-b}
\end{subfigure}
\caption{Constant job arrival with 100 jobs using full clone. \vspace{3mm}}
\label{mixed-full-periodic}
\end{figure*}



\begin{table}[]
\caption{The overheads incurred by Multiverse framework for VM provisioning and allocation.}
\label{tbl:overheads}
\footnotesize
\begin{tabular}{|l|l|}
\hline
\textbf{Overhead Type}         & \textbf{Description}                                                                                                             \\ \hline
\textbf{schedule\_clone}       & \begin{tabular}[c]{@{}l@{}}Time taken to start the clone script \\ by VM\_Launch daemon.\end{tabular}                            \\ \hline
\textbf{get\_host}             & \begin{tabular}[c]{@{}l@{}}Time taken to get a compatible host \\ from load balancer.\end{tabular}                               \\ \hline
\textbf{network\_configuration} & \begin{tabular}[c]{@{}l@{}}Time taken to configure and customize\\ the network.\end{tabular}                                     \\ \hline
\textbf{slurmd\_customization}  & \begin{tabular}[c]{@{}l@{}}Time taken to copy the Slurm config files \\ and restart the slurmd node daemons.\end{tabular}       \\ \hline
\textbf{slurm\_restart}        & \begin{tabular}[c]{@{}l@{}}Time taken by VM Launch daemon to\\ restart slurmctld after VM is ready.\end{tabular}                 \\ \hline
\textbf{slurm\_schedule}       & \begin{tabular}[c]{@{}l@{}}Time taken to assign the job to the VM\\ after restarting the controller.\end{tabular} \\ \hline
\end{tabular}
\end{table}
\subsection{Results and Analysis}
\label{sec:results-explained}
Figure~\ref{mixed-full-50-c} and \ref{mixed-instant-50-c} shows the breakdown {of} the overall job completion {time} for 50 jobs using full and instant clone respectively. It can be seen that instant clone is extremely fast in provisioning VMs with an average cloning time of 10s. On the other hand, full clone takes about 150s on average with a maximum clone time of 450s in some cases. This is because, full clones do not share any resources with the parent VM {and require} a lot more time for disk provisioning. Instant clone, on the other hand, are forked off of the parent VM and hence are very fast. However, from Figure \ref{mixed-instant-50-b} which shows the break down of individual overheads, the network configuration overhead is very high for instant clones. \emph{Since they share the same network config as the parent VM after cloning, the network has to be re-configured}. However, instant clones (36s on average) are still 7.2$\times$ faster compared to full clones (260s on average), with respect to overall VM provisioning time. 

The overhead to restart the \textit{Slurm} controller by the VM launch daemon is in the order of 20s. Majority of this time is spent within \textit{Slurm} to complete the restart after being initiated by the daemon. Since a lot of jobs are running on the system, the scheduler spends significant amount of time for the restart. 
The other overheads apart from network\_config and slurmctld\_restart are minimal and very similar for both clone types (shown in Figure~\ref{mixed-full-50-b} and \ref{mixed-instant-50-b}). Figure~\ref{job_completion} plots the job running time using both instant and full clone. The running times are fairly similar with a few variations. We can infer that running time does not get affected due to type of clone used. 

Figure \ref{mixed-50-full-periodic} shows the breakdown of job completion and other overheads for a {constant} inter-arrival (10s) between jobs. The total cloning time is within 75s, and the overall provisioning time including other overheads is within 140s for all the jobs. In this case, full clone performs very similar to instant clone in terms of overall job completion time. This is because, the number of concurrent clone operations handled by vSphere is significantly reduced. vSphere can handle up to 200 concurrent {instant} clones but it incurs higher latency for concurrent full clones. However, the overall provisioning time using instant clone (36s on average) is still 2.5$\times$ faster than full clone (87s on average). Note that, the job running times are much lower for constant arrival when compared to bursty arrival. This is due to the fact that, the cluster is not 100\% utilized for a constant arrival, which consequently leads to lesser interference among VMs. \\ 
\subsubsection{\textbf{Scalability using CPU Over-commitment}} We conduct another set of experiments where we use 2x over-commitment of CPU in the cluster (i.e), the cluster will be running jobs with total vCPUs equal to twice as {many as} available CPUs. We use a the same Poisson based arrival sequence, but for 100 jobs. Figure~\ref{mixed-100-c} plots the breakdown of job completion time for instant and full clone. It can be seen that, instant clone scales well for 100 jobs as the cloning time is well within 15s (Figure~\ref{mixed-instant-c}) for all the jobs. For the last few jobs from 86 to 91, the overheads are higher because the cluster is already full with no more available vCPUs to allocate for VMs. This is shown in  Figure~\ref{mixed-instant-b}, that the get\_host time for these jobs are very high. On the other hand, for full clone, the time {taken} to clone is very large (shown in Figure~\ref{mixed-full-c}). This is because concurrent full clones are very slow to handle. As described in Section~\ref{sec:modeling}, we use a rate-limiter of 15 clones per minute. This can be seen in Figure~\ref{mixed-full-b}, where the schedule\_clone increases in multiples of 15s. The performance degradation due to clone overheads is very large for jobs starting from 51, because more and more jobs are queued in vSphere to be cloned. Note that, for instant clone, some jobs (41, 42 etc shown in Figure~\ref{mixed-instant-b}) take a longer time to complete than the average. This is because, the 2x CPU over-commitment causes a CPU pressure on the physical hosts as we have more allocated virtual CPUs than available CPUs.
We repeat the experiment for same job configuration but using a {constant} inter-arrival time of jobs for full clone. As seen in Figure~\ref{mixed-full-periodic-c}, the cloning time is much faster compared to workload-2 and the overall job completion is very similar to instant clone. As stated earlier, this is because increasing the inter-arrival time between jobs leads to fewer concurrent clones executing in the cluster. Also there are no straggler jobs towards the end as in instant clone because, all the 100 jobs are equally spaced out in the cluster and do not run concurrently. We can conclude that, instant clone is best suited in case of bursty job arrivals as opposed to full clone, which would be suitable for a constant job arrival rate. Furthermore, we can build an mixed system that can use a combination of instant and full clones, depending on the difference in job arrival rate over time.


\subsubsection{\textbf{Cluster Utilization and Throughput}} 
\begin{wrapfigure}{l}{0.25\textwidth}
    \centering
    \includegraphics[width=.95\linewidth]{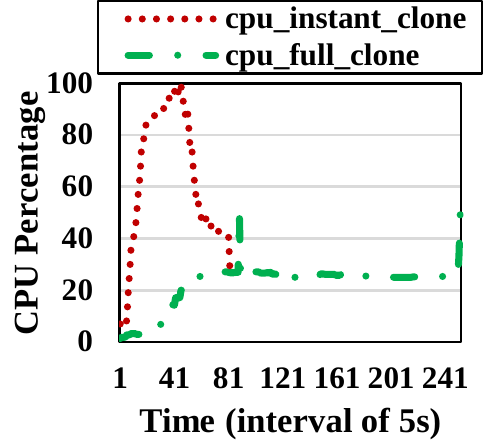}
\caption{Average CPU utilization.}
\label{util}
\end{wrapfigure}

Figure~\ref{util} plots the CPU utilization for workload-2 using both full and instant clones. The utilization numbers are collected periodically every 10s for the entire workload execution time. It can be seen that the average CPU utilization for instant clone, initially is 60\%; but, starting from time-step 21, it is very high ranging from 80-100\%. This clearly indicates that all the 100 jobs are scheduled as soon as they arrive and are executing concurrently in the cluster. On the other hand for full clone, the maximum cluster CPU utilization never goes beyond 50\%. This is because most of the jobs spend a lot of time to get a cloned VM before they can start executing. This reduces the number of concurrent jobs in the system. Further, the system throughput using instant clone is 1.5$\times$ better than using full clone. This is because, the total time taken for job completion is 581s (end of dotted-red line) for instant clone, when compared to 868s (end of dashed-green line) for full clone.
\subsubsection{\textbf{Comparison with bare-metal deployments}} 
\begin{figure}[hbpt]
\centering
\includegraphics[width=0.99\linewidth]{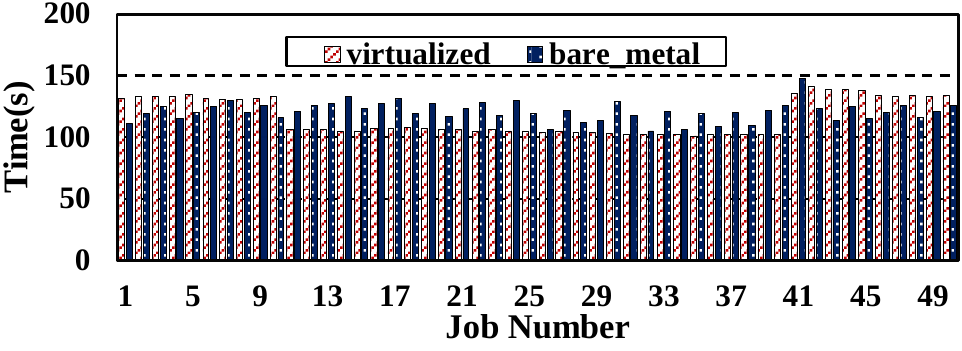}
\caption{Job running times of virtualized deployment compared to bare-metal deployment.}
\label{bare_metal}
\end{figure}

To ensure that, the job performance is not affected due to virtualization overheads, we conduct another set of experiments by executing the jobs in a bare-metal cluster managed by \textit{Slurm}.  Figure~\ref{bare_metal} shows the comparison of job running times for both bare-metal and virtualized deployments. The running/execution times are fairly similar without significant variations. Hence, we can conclude that virtualization can deliver near-native bare-metal performance for jobs.\vspace{3mm}  
\subsection{Overheads of Multiverse}
We explain in detail the overheads incurred with respect to the different components of multiverse. \\
\subsubsection{\textbf{Clone Overheads}} Despite instant clone being faster than full clone, the network overhead incurred by instant clone is significantly high. This is because, they replicate the same network configuration as the parent VM. We need to manually set the IP address (or use DHCP) and reconfigure the network of the cloned VM. This incurs about 10-20s of the total VM start-up time. 
\subsubsection{\textbf{Controller Restart Overheads}} As explained in Section~\ref{sec:results-explained}, restarting the slurm controller takes around 20s on average. This overhead can be significantly minimized by (i) increasing the size (CPU and memory allocations) for the slurm controller VM, and (ii) multiple slurm controllers can be used in parallel to share the load of managing the slave VMs. We also mention in Section~\ref{sec:overheads} that restarting the controller can be avoided if we use other HPC schedulers like PBS or Torque. 
\subsubsection{\textbf{Job Concurrency Overheads}} Our Workload-1 does not have any interference from jobs because, it consists of stream of 50 jobs that can entirely fit in the physical capacity (CPU and memory) of the 220-core cluster. However, in Workload-2 the overall job requires 2x of available physical capacity (2x over-commitment). Therefore there is certain degree of performance degradation in terms of job running times (Figure~\ref{mixed-instant-c} vs Figure~\ref{mixed-instant-50-c}). We can further characterize the tolerance of over-commitment ratio with respect to job performance, but that is beyond the scope of this paper. For constant job-arrival, there is no impact of interference because the jobs are equally spaced without any contention for resources. 

%% file: 7-related.tex
\noindent{\textbf{Dynamic VM Provisioning for HPC:}}
With the prevalence of virtualization into the HPC community, some prior works have attempted to integrate a dynamic VM provisioning model using HPC scheduler like Slurm and Torque~\cite{formosa3,meier2016dynamic}. However, these frameworks are neither robust nor eliminate manual user intervention. The most relevant work to \textit{Multiverse} is proposed by Zhange et al~\cite{vslurm} called v-slurm. However, they do not characterize the other overheads apart from VM provisioning. We provide a detailed characterization of all overheads associated with the \textit{Multiverse} framework. Moreover, \textit{Multiverse} is the first work to employ instant clone based rapid dynamic VM provisioning in a HPC environment. 
\\
{\textbf{Agile VM Provisioning:}} There are several prior works which have proposed quick and agile VM provisioning mechanisms. Some of them require live VMs~\cite{lagar2009snowflock}, while the rest try to minimize the VM disk size~\cite{razavi2015prebaked}. However, none of these have been adopted by a majority of  mainstream HPC private clusters. There has recently been an increased interest in running containers such as Docker inside VMs~\cite{ZhangLP16}. One of the major benefits of such an approach is fast environment startup -- in the order of seconds. Containers, however, have dependency on their hosting OS, due {to} their process-based nature. This makes container migration difficult. On the other hand, instant clones used in \textit{Multiverse} are very fast, and comparable to container based provisioning times.\\
\textbf{HPC in public cloud:} There has been significant advancements in HPC provisioning by the union of cloud system stack along with HPC, enabling IaaS-based provisioning of HPC infrastructures. In this context, many dynamic VM provisioning schemes have been proposed using Microsoft Azure, AWS EC2, etc,~\cite{garg2014sla}. However, HPC clusters are largely hosted in private datacenters for security, tractability, and fault tolerance. Our work primarily focuses on mitigating the bottlenecks of virtualized HPC in a private setting. Also,  some of the ideas proposed in the context of public cloud,~\cite{public,spock,knots,phoenix} can also be leveraged by \textit{Multiverse}, as our framework is a generic implementation by setting up a platform for further enhancements.

%% file: 8-conclusion.tex

In this paper, we identify several challenges in developing a dynamic VM provisioning framework for virtualized HPC clusters. We design and implement  \textit{Multiverse}, which integrates an HPC scheduler with a VM orchestrator, to dynamically spawn  VMs for incoming jobs in a virtualized HPC cluster. This enables more flexible and cluster utilization aware VM provisioning. We further explore the potential of using instant cloning compared to full cloning in terms of VM provisioning overhead, resource utilization and cluster throughput. 
Experimental results with HPC workloads indicate that instant cloning on an average is  $2.5\times$ - $7.2\times$  faster than full cloning in terms of VM provisioning time. Further, instant cloning improves cluster utilization by up to 40\% and cluster throughput by up to $1.5\times$, when compared to full clones for bursty job arrivals. On the other hand, full cloning is comparatively better for constant job arrivals with large inter-arrival times.
In our future work, we plan to compare instant clone based VM provisioning against a container-based provisioning. Towards this, we plan to integrate a docker hypervisor with the Slurm scheduler and analyze the job completion times with our \textit{Multiverse} framework.

%% file: 9-acknowledgment.tex
We are indebted to Na Zhang, Anup Sarma and Cyan Mishra for their insightful comments on several drafts of this paper. This research was partially supported by NSF grants \#1931531, \#1629129, \#1763681, \#1629915, \#1908793, \#1526750 and we thank NSF Chameleon Cloud project CH-819640 for their generous compute grant. We also thank Mohan Potheri for providing us with a compute cluster from VMware to conduct all the experiments. 